\documentstyle[preprint,aps]{revtex}
\begin{document}
\draft
\title{Quantum Solitons in the Calogero-Sutherland Model}
\author{R. K. Bhaduri and Akira Suzuki \footnotemark[1]}
\address{Department of Physics and Astronomy, McMaster University,}
\address{Hamilton, Ontario, Canada L8S 4M1}
\footnotetext[1]{Permanent address: Department of Physics, Science
University of Tokyo, Tokyo, Japan}
\maketitle
\begin{abstract}
We show that the single quasi-particle Schr\"odinger equation for
a certain form of one-body potential yields a stationary one soliton
solution.
The one-body potential is assumed to arise from the self-interacting
charge distribution with the singular kernel of the Calogero-Sutherland
model.

The quasi-particle has negative or positive charge for negative or
positive coupling constant of the interaction.
The magnitude of the charge is unity only for the semion.
It is also pointed out that for repulsive coupling, our equation is
mathematically the same as the steady-state Smoluchowski equation of
Dyson's Coulomb gas model.
\end{abstract}

Recently, interesting work has been done\cite{suth94,poly95} on
classical solitons in the one-dimensional many-body system with
inverse-square interaction.
It has been found that in the $\hbar\rightarrow0$ limit, particle
excitations from the quantum ground-state may be regarded as solitons.
Two different approaches have been used to investigate this problem.
Sutherland and Campbell\cite{suth94} have examined Newton's equation of
motion for finite systems, and then taken the continuum (thermodynamic)
limit for the infinite system.
In the second approach, Polychronakos\cite{poly95} has exploited the
quantum collective field formulation of Andric {\it et al.}\cite{and83}
to obtain the one-soliton solution in the $\hbar\rightarrow0$ continuum
limit.
In this paper, on the other hand, we enquire under what conditions
a particle obeying the Schr\"odinger equation will give rise to
stationary solitonic solutions in the Calogero-Sutherland model
\cite{calo69,suth71a,suth71b,suth72}(CSM).
We find that the one-body Schr\"odinger equation may be cast in the form
of a solvable nonlinear diffusion equation when a certain ansatz for
the one-body potential is made.
The one-soliton solution of this diffusion equation, first proposed
by Satsuma and Mimura\cite{satsu85}, is of the same form as found by
Polychronakos\cite{poly95}.
We find that soliton solutions may exist for both repulsive or
attractive interactions, and these carry positive or negative
fractional charge inversely proportional to the strength of the
interaction.

We begin with the Schr\"odinger equation for a particle propagating
in a one-body potential $W(x)$.
The treatment is completely general at this stage, and follows the
Feynman Lectures\cite{feyn65} on physics.
The ansatz for $W(x)$ will be made later to obtain the soliton solutions.
Denoting the one-particle wave-function by $\psi(x,t)$, we have
\begin{equation}
-{\hbar^{2} \over 2m}{\partial^{2}\psi(x,t) \over {\partial x^{2}}} \;
+ \; W(x)\psi(x,t) \; = \;
i\hbar{\partial\psi(x,t)\over{\partial t}} \;\;.
\label{scheq}
\end{equation}
Henceforth we shall denote the partial derivatives with respect to $x$
by a prime and with respect to $t$ by a dot.
It is useful to multiply Eq.~(\ref{scheq}) by $\psi^{*}(x,t)$,
\begin{equation}
-{\hbar^{2} \over 2m}\psi^{*}\psi'' \; + \; \psi^{*}W(x)\psi \;
= \; i\hbar\psi^{*}\dot\psi \;\;,
\label{hden1}
\end{equation}
and take the complex conjugate:
\begin{equation}
-{\hbar^{2} \over 2m}\psi{\psi^{*}}'' \; + \; \psi W(x)\psi^{*}
\; = \; -i\hbar\psi\dot\psi^{*} \;\;.
\label{hden2}
\end{equation}

On subtracting Eq.~(\ref{hden2}) from Eq.~(\ref{hden1}), we get
\begin{equation}
-{\hbar^{2}\over 2m}(\psi^{*}\psi''-{\psi^{*}}''\psi)
\; = \; i\hbar(\psi^{*}\dot\psi+\psi\dot\psi^{*}) \;\;.
\label{hden32}
\end{equation}
 We now set, quite generally,
\begin{equation}
\psi(x,t) \; = \; {\sqrt{\rho(x,t)}}\;\;{\rm e}^{i\theta(x,t)} \;\;,
\label{defpsi}
\end{equation}
where $\rho$ and $\theta$ are real.
Then Eq.~(\ref{hden32}) reduces to the continuity equation for the
``charge'' density $\rho$:
\begin{equation}
\dot\rho \; + \; {\hbar\over m}{\partial\over{\partial x}}(\rho\theta')
\; = \; 0\;\;.
\label{conteq}
\end{equation}
Thus it is legitimate to rgard the velocity of the ``fluid'' to be
\begin{equation}
v(x,t) \; = \; {\hbar\over m}\theta(x,t) \;\;.
\label{velocity}
\end{equation}
The energy density equation is obtained by adding Eqs.~(\ref{hden1})
and (\ref{hden2}), and dividing by 2:
\begin{equation}
\tau(x,t) \; + \; W(x)\rho \; = \; - \hbar\rho\dot\theta \;\;,
\label{enden}
\end{equation}
where the kinetic energy density is
\begin{eqnarray}
\tau(x,t) \; & = & \; -{\hbar^{2} \over 4m}(\psi^{*}\psi''
                                   +{\psi^{*}}''\psi) \nonumber \\
             & = & \; {\hbar^{2} \over 2m}
               \left({1\over4}{(\rho')^{2} \over \rho}
              -{1\over2}\rho''+\rho(\theta')^{2}\right) \;\;.
\label{kinetic}
\end{eqnarray}
We note, however, that an alternate form of kinetic energy density is
given by
\begin{eqnarray}
\tau_{1}(x,t) \; & = & \; {\hbar^{2}\over 2m}({\psi^{*}}'\psi') \nonumber \\
                 & = & \; {\hbar^{2}\over 2m}\left({1\over 4}
                 {(\rho')^{2} \over \rho}+\rho(\theta')^{2}\right) \;\;.
\label{kinetic1}
\end{eqnarray}
The two expressions for the kinetic energy density differ by
a perfect differential $(\hbar^{2}/4m)\rho''$ which vanishes
on integration over space provided $\rho'=0$ at infinity.
We exploit this ambiguity to our advantage, and choose the kinetic
energy density $\tilde\tau$ to be the mean of $\tau$ and $\tau_{1}$, i.e,
\begin{eqnarray}
\tilde\tau(x,t) \; & = & \;
                 {1\over2}\left\{\tau(x,t)+\tau_{1}(x,t)\right\} \nonumber \\
                 & = & \; {\hbar^{2}\over 2m}\left({1\over 4}
{(\rho')^{2}\over\rho}-{1\over4}\rho''+\rho(\theta')^{2}\right) \;\;.
\label{kinetic2}
\end{eqnarray}
In the energy density Eq.~(\ref{enden}), $\tau(x,t)$ is now replaced by
$\tilde\tau(x,t)$, and we obtain
\begin{equation}
{\hbar^{2}\over 2m}\left({1\over4}{(\rho')^{2}\over\rho}-{1\over4}\rho''
+\rho(\theta')^{2}\right) \; + \; W(x)\rho \; = \; - \hbar\rho\dot\theta \;\;.
\label{modtau}
\end{equation}
Dividing through by $\rho$, and taking the derivative with respect to $x$,
we get
\begin{equation}
m\left({\partial\over{\partial t}}+v{\partial\over{\partial x}}\right)v \;
= \; {\hbar^{2}\over {8m}}{\partial^{2}\over{\partial x^{2}}}
\left({\rho'\over\rho}\right) \; - \; {\partial W\over{\partial x}}\;\;.
\label{moving}
\end{equation}
In the ``co-moving'' frame of the fluid, the left-hand side is the
total derivative $m(dv/dt)$, and vanishes for uniform propagation
velocity.
We then obtain the steady-state equation in this frame to be
\begin{equation}
{\hbar^{2}\over {8m}}{\partial\over{\partial x}}\left({\rho'\over
\rho}\right) \; - \; W(x) \; = \; 0 \;\;,
\label{oureq}
\end{equation}
where the constant of integration is taken to be zero.
Until now, our treatment has been completely general, except for
the special choice ${\tilde\tau}(x,t)$ of the kinetic energy density.
We now proceed to investigate the form of $W(x)$ for which we may
obtain soliton solutions.

We have in mind a one-dimensional many-particle system interacting
pairwise with an inverse-square potential\cite{calo69,suth71a}.
The hamiltonian for $N$-particles is given by
\begin{equation}
H \; = \; -{\hbar^{2}\over{2m}}\sum^{N}_{i=1}{\partial^{2}\over{\partial
x_{i}^{2}}} \;
+ \; {\hbar^{2}\over m}g\sum_{i<j}{1\over {(x_{i}-x_{j})^{2}}} \;
+ \; {1\over 2}m\omega^{2}\sum_{i=1}^{N}x_{i}^{2} \;\;,
\label{hamiltonian}
\end{equation}
where $g$ is dimensionless and may take any value $\ge -1/2$.
In the range $-1/2\le g \le 0$ it is related to the statistical
parameter $\alpha$ of exclusion statistics\cite{hal91} by
the relation $g=\alpha(\alpha-1)$.
The particles are confined by a harmonic potential, and the
thermodynamic limit is taken by letting $\omega\rightarrow 0$ as
$N^{-1}$.
Generally we have the $N\rightarrow\infty$ limit in mind with
a vanishing harmonic confinement.
For such a system, we make the ansatz that a quasi-particle
excitation obeying Eq.~(\ref{oureq}) experiences a potential
$W(x)$ given by (P is the principal value of the integral)
\begin{equation}
W(x) \; = \; -{\hbar^{2}\over m}g{\partial\over{\partial x}}
{\rm P}\int_{-\infty}^{\infty}{\rho(y)dy\over{x-y}} \;\;,
\label{ansatz}
\end{equation}
with a finite normalization corresponding to a charge $q$:
\begin{equation}
\int_{-\infty}^{\infty}\rho(y)dy \; = \; q \;\;.
\label{norm}
\end{equation}
We emphasize that Eq.~(\ref{ansatz}) is {\em not} a mean-field potential,
but is a potential that arises from the {\em self-interaction} of the
localized charge-distribution of the quasi-particle.
With this ansatz, Eq.~(\ref{oureq}) takes the form
\begin{equation}
{\partial\over{\partial x}}\left[{1\over 8}\left({\rho'\over\rho}\right)
+ g{\rm P}\int_{-\infty}^{\infty}{\rho(y)dy\over{x-y}}\right] \; = \; 0 \;\;.
\label{rhoeq1}
\end{equation}
This is obviously satisfied if $\rho(y)$ is a solution of
\begin{equation}
{\rho'\over\rho} \;
= \; -8g{\rm P}\int_{-\infty}^{\infty}{\rho(y)dy\over{x-y}} \;\;.
\label{rhoeq2}
\end{equation}
Note that Eq.~(\ref{rhoeq1}) may then be rewritten in the form
\begin{equation}
\rho'' \; + \; 8g{\partial\over{\partial x}}
\left[{\rm P}\int_{-\infty}^{\infty}{\rho(y)dy\over{x-y}}\rho(x)\right]
\; = \; 0 \;\;.
\label{rhoeq3}
\end{equation}
Formally, this equation has the same form as the steady-state
Coulomb gas model of Dyson\cite{dyson62,suth72} for $g>0$.
In the diffusion problem, it is the steady-state Smoluchowski equation
with a singular kernel, which describes the Brownian motion of
a particle immersed in a fluid, with a friction-limited velocity.
A clear description of this equation starting from the Langevin
equation is given by Andersen and Oppenheim\cite{and63}.
We stress that although formally our Eq.~(\ref{rhoeq3}) is exactly
of the same form as the diffusion equation for Brownian particles,
the physics is quite different.
Our system is conservative, while the latter is not, being subject
to frictional forces.

Following the work of Satsuma and Mimura\cite{satsu85}, we can write
down the one-soliton stationary solution of Eq.~(\ref{rhoeq3}).
It is straight-forward to check, through a contour integration, that
Eq.~(\ref{rhoeq2}) is satisfied by
\begin{equation}
\rho(x) = {c\over{4\pi g}}\;{1\over{x^{2}+c^{2}}} \;\;,
\label{solution1}
\end{equation}
where $c$ is an arbitrary positive parameter.
It follows from Eq.~(\ref{solution1}) that the charge carried by the
soliton is
\begin{equation}
\int_{-\infty}^{\infty}\rho(x)dx \; = \; {1\over{4g}} \;\;.
\label{charge}
\end{equation}
This shows that the charge carried by the soliton is inversely
proportional to $g$, and is negative unity for $\alpha=1/2$ (semion).

Following Ref.~\cite{suth72}, we may also choose the appropriate
kernel for $W(x)$ in Eq.~(\ref{ansatz}) to obtain the soliton
solutions for the Sutherland Hamiltonian\cite{suth71b} on a circle.
The many-particle system on a circle of circumference $L$ is
given by
\begin{equation}
H \; = \; -{\hbar^{2}\over{2m}}\sum_{i}{\partial^{2}\over{\partial x_{i}^{2}}}
\; + \; {\hbar^{2}\over m}g\left({\pi\over L}\right)^{2}\sum_{i<j}
\left(\sin\left[{\pi(x_{i}-x_{j})\over L}\right]\right)^{-2} \;\;.
\label{suthH}
\end{equation}
In this case, we choose
\begin{equation}
W(x) \;=\; -{\hbar^{2}\over m}g{\pi\over L}{\partial\over{\partial x}}
{\rm P}\int_{-L/2}^{L/2}\cot\left[{\pi(x-y)\over L}\right]\rho(y)dy \;\;,
\label{suthW}
\end{equation}
where $\rho(y)$ is taken to be a periodic solution, $\rho(y)=\rho(y+L)$.
The equation satisfied by $\rho$ is
\begin{equation}
\rho''\;+\;8g{\pi\over L}{\partial\over{\partial x}}
\left\{{\rm P}\int_{-L/2}^{L/2}\cot\left[{\pi(x-y)\over L}\right]
\rho(y)dy\;\rho(x)\right\} \;=\; 0 \;\;.
\label{sutheq}
\end{equation}
Following Ref.~\cite{suth72}, the solution is
\begin{equation}
\rho(x) \;=\; {n\over{4gL}}\;{{\rm sinh}\phi \over
{\cos[(2n\pi/L)x] + {\rm cosh}\phi}} \;\;,
\label{suthsol}
\end{equation}
where $n$ is an integer and $\phi$ a positive constant.
The normalization  of the density is independent of $\phi$, and is
given by
\begin{equation}
q \;=\; \int_{-L/2}^{L/2}\rho(x)dx \;=\; {n\over{4g}} \;\;.
\label{suthnorm}
\end{equation}
Polychronakos \cite{poly95} obtained the periodic form of (\ref{suthsol})
by a superposition of the Lorentzians given by Eq.~(\ref{solution1}).
This gives
\begin{eqnarray}
\rho(x) & = & {c\over{4\pi g}}\sum_{j=-\infty}^{\infty}
              {1\over{(x+j\lambda)^{2}+c^{2}}} \nonumber \\
        & = & {1\over{4\lambda g}}\;{\sinh(2\pi c/\lambda)\over
              {\cosh(2\pi c/\lambda)-\cos(2\pi x/\lambda)}} \;\;.
\label{solutionpoly}
\end{eqnarray}
In general, this is not a solution of the original nonlinear
Eq.~(\ref{rhoeq3}) unless a special condition is satisfied\cite{poly95}.
However, Eq.~(\ref{solutionpoly}) is always a solution of
Eq.~(\ref{sutheq}) corresponding to the Sutherland Hamiltonian on
a circle.
Our solution (\ref{suthsol}) may be recovered from
Eq.~(\ref{solutionpoly}) by shifting $x\rightarrow(x+\lambda/2)$,
and by setting $n\lambda=L$.

In concluding, it should be pointed out that our Eq.~(\ref{oureq})
for the charge density $\rho(x)$ was derived by assuming a special form
$\tilde\tau(x)$ of the kinetic energy density (see Eq.~(\ref{kinetic2})).
Alternatively, we could have directly substituted $\psi=\sqrt\rho\;
{\rm e}^{i\theta}$ in the one-particle Schr\"odinger equation
(\ref{scheq}), and added an additional term $(\hbar^{2}/8m)(\rho''/\rho)$
to $W(x)$ to obtain the same result.
The main point of the paper is that it is possible, under certain
assumptions, to obtain the same type of mathematical solutions in
a quantum frame-work as found in Ref.~\cite{poly95}, in which
the continuum limit with $\hbar\rightarrow 0$ was taken.

This research was supported by NSERC of Canada.
We would like to thank M.V.Murthy for useful discussions.
One of the authors (A.S) would like to thank the theory group of
Department of Physics and Astronomy of McMaster University for
hospitality during the period of his stay in the summer of 1995.

\end{document}